\documentclass[11pt,twoside]{article}  % Leave intact
\usepackage{asp2006}
\usepackage{adassconf}

\def\teff{${\rm T}_{\rm eff}$}
\def\logteff{$\log$\,(${\rm T}_{\rm eff}$)}
\def\logg{$\log g$}

\begin{document}   % Leave intact

%-----------------------------------------------------------------------
%                           Paper ID Code
%-----------------------------------------------------------------------
% Enter the proper paper identification code.  The ID code for your paper 
% is the session number associated with your presentation as published 
% in the official conference proceedings.  You can find this number by 
% locating your abstract in the printed proceedings that you received 
% at the meeting, or on-line at the conference web site.
%
% This identifier will not appear in your paper; however, it allows different
% papers in the proceedings to cross-reference each other.  Note that
% you should only have one \paperID, and it should not include a
% trailing period.

\paperID{O9.3}

%-----------------------------------------------------------------------
%                           Paper Title 
%-----------------------------------------------------------------------
% Enter the title of the paper.
%
% EXAMPLE: \title{A Breakthrough in Astronomical Software Development}

\title{A method for exploiting domain information in astrophysical parameter estimation}
       
%-----------------------------------------------------------------------
%          Short Title & Author list for page headers
%-----------------------------------------------------------------------
% Please supply the author list and the title (abbreviated if necessary) as 
% arguments to \markboth.
%
% The author last names for the page header must appear in one of 
% these formats:
%
% EXAMPLES:
%     LASTNAME
%     LASTNAME1 and LASTNAME2
%     LASTNAME1, LASTNAME2, and LASTNAME3
%     LASTNAME et al.
%
% Use the "et al." form in the case of four or more authors.
%
% If the title is too long to fit in the header, shorten it: 
%
% EXAMPLE: change
%    Rapid Development for Distributed Computing, with Implications for the Virtual Observatory
% to:
%    Rapid Development for Distributed Computing

\markboth{Bailer-Jones}{Exploiting domain information in parameter estimation}

%-----------------------------------------------------------------------
%                         Authors of Paper
%-----------------------------------------------------------------------
% Enter the authors followed by their affiliations.  The \author and
% \affil commands may appear multiple times as necessary.  List each
% author by giving the first name or initials first followed by the
% last name. Do not include street addresses and postal codes, but 
% do include the country name or abbreviation. 
%
% If the list of authors is lengthy and there are several institutional 
% affiliations, you can save space by using the \altaffilmark and \altaffiltext 
% commands in place of the \affil command.

\author{C.A.L.\ Bailer-Jones}
\affil{Max Planck Institute for Astronomy, Heidelberg, Germany}

% Notice that some of these authors have alternate affiliations, which
% are identified by the \altaffilmark after each name.  The actual alternate
% affiliation information is typeset in footnotes at the bottom of the
% first page, and the text itself is specified in \altaffiltext commands.
% There is a separate \altaffiltext for each alternate affiliation
% indicated above.

%-----------------------------------------------------------------------
%                        Contact Information
%-----------------------------------------------------------------------
% This information will not appear in the paper but will be used by
% the editors in case you need to be contacted concerning your
% submission.  Enter your name as the contact along with your email
% address.

\contact{Coryn Bailer-Jones}
\email{calj@mpia.de}

%-----------------------------------------------------------------------
%                     Author Index Specification
%-----------------------------------------------------------------------
% Specify how each author name should appear in the author index.  The 
% \paindex{ } should be used to indicate the primary author, and the
% \aindex for all other co-authors.  You MUST use the following syntax: 
%
%    \aindex{LASTNAME, F.~M.}
% 
% where F is the first initial and M is the second initial (if used). Please 
% ensure that there are no extraneous spaces anywhere within the command 
% argument. This guarantees that authors that appear in multiple papers
% will appear only once in the author index. Authors must be listed in the order
% of the \paindex and \aindex commmands.

\paindex{Bailer-Jones, C.~A.~L.}

%-----------------------------------------------------------------------
%                       Subject Index keywords
%-----------------------------------------------------------------------
% Enter up to 6 keywords that are relevant to the topic of your paper.  These 
% will NOT be printed as part of your paper; however, they will guide the creation 
% of the subject index for the proceedings.  Please use entries from the
% standard list where possible, which can be found in the index for the 
% ADASS XVI proceedings. Separate topics from sub-topics with an exclamation 
% point (!). 


% We reset the footnote counter for the hyperlink since it does not
% appear to recognize the previous 3 footnotes generated from the
% altaffilmarks.  

%\setcounter{footnote}{3}

%-----------------------------------------------------------------------
%                              Abstract
%-----------------------------------------------------------------------
% Type abstract in the space below.  Consult the User Guide and Latex
% Information file for a list of supported macros (e.g. for typesetting 
% special symbols). Do not leave a blank line between \begin{abstract} 
% and the start of your text.

\begin{abstract}          % Leave intact
  I outline a method for estimating astrophysical parameters (APs) from
  multidimensional data.  It is a supervised method based on matching observed
  data (e.g.\ a spectrum) to a grid of pre-labelled templates. However, unlike
  standard machine learning methods such as ANNs, SVMs or k-nn, this algorithm
  {\em explicitly} uses domain information to better weight each data
  dimension in the estimation.  Specifically, it uses the sensitivity of each
  measured variable to each AP to perform a local, iterative interpolation of
  the grid. It avoids both the non-uniqueness problem of global regression
  as well as the grid resolution limitation of nearest neighbours.
\end{abstract}

%-----------------------------------------------------------------------
%                             Main Body
%-----------------------------------------------------------------------
% Place the text for the main body of the paper here.  You should use
% the \section command to label the various sections; use of
% \subsection is optional.  Significant words in section titles should
% be capitalized.  Sections and subsections will be numbered
% automatically. 

\section{Introduction}

Consider the problem of estimating the astrophysical
parameters (APs) of a star from its spectrum using a grid of pre-labelled
spectra. Let ${\vec p} = \{p_i\}$, $i=1 \ldots I$ be the data vector (spectrum
or multiband photometry) and ${\vec \phi} = \{\phi_j\}$, $j=1 \ldots J$ be the
AP vector (e.g.\ \teff, \logg, etc.).  Standard approaches involve performing
a global regression on the grid to infer the mapping $\vec \phi = g(\vec p)$,
using, for example, a artificial neural network (ANN) (e.g.\ Bailer-Jones et
al.\ 1998) or a support vector machine (SVM) (e.g.\ Tsalmantza et al.\ 2006).

\begin{figure}[t]
\epsscale{0.80}
\plotone{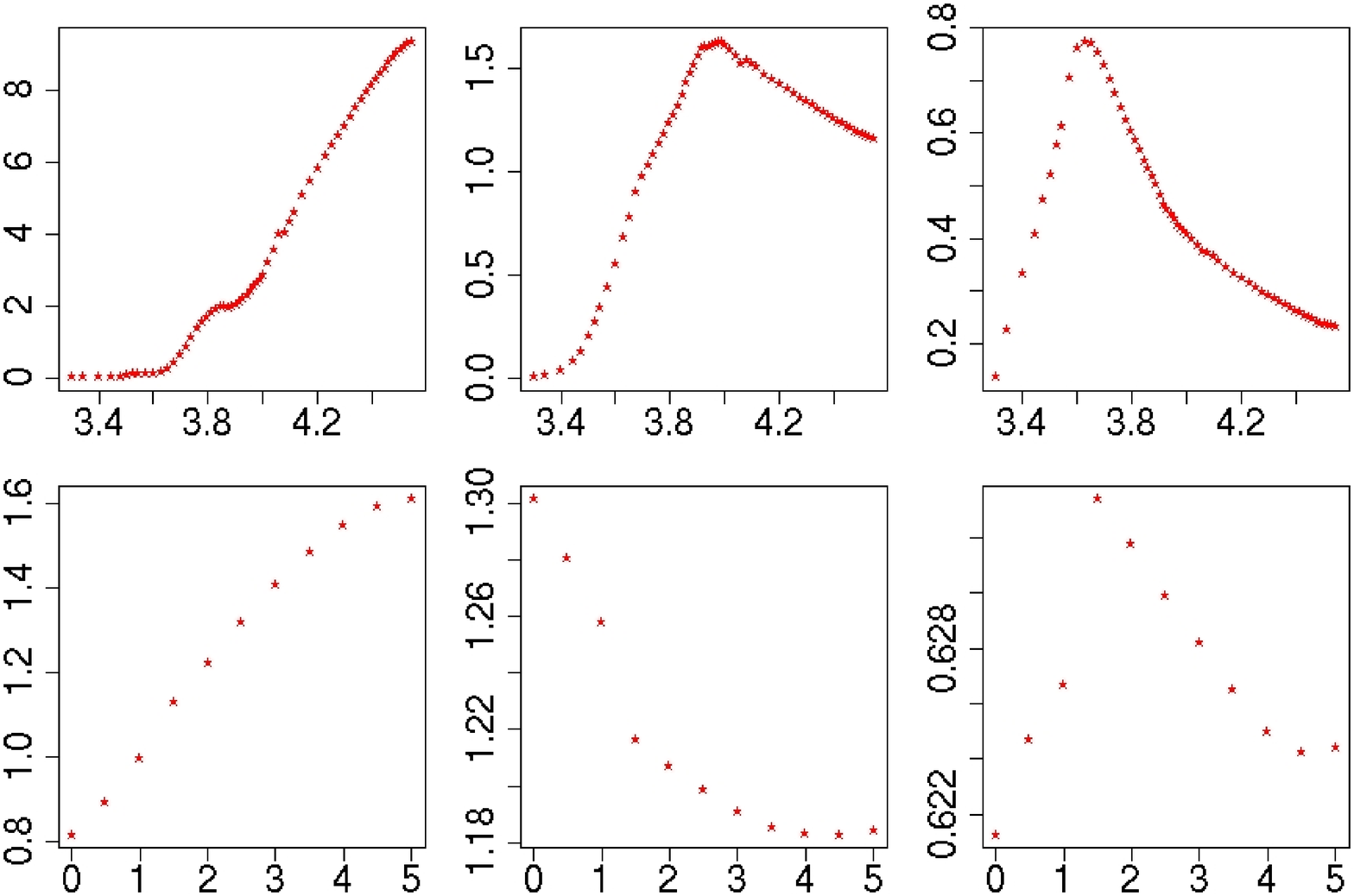}
\caption{Variation of photon counts with \logteff\ (top) and \logg\ (bottom) in three filters (bands). A $J$-dimensional fit to each band (independently) is a forward model. This is a true function (unlike a fit to the inverse).} \label{formod}
\end{figure}

Although these methods meet with reasonable success, they have problems when
it comes to estimating multiple APs, in particular if some APs have a
relatively weak signature (as is the case with \logg\ and [Fe/H]: compare the
vertical scales in Fig.~\ref{formod}). To overcome this we should weight the
variables according to their sensitivity with respect to the APs of interest.
In principle, ANNs and SVMs implicitly learn this weighting from the data, but
this is difficult with many noisy variables. Furthermore, a global regression
approach is strictly flawed, because while the photon counts in a band varies
uniquely with the APs, the converse is not true (Fig.~\ref{formod}).  The
global regression is trying to solve an inverse problem and the lack of
uniqueness could lead to a poor fit.  This degeneracy problem is exacerbated
at low spectral resolution and by noise.

\section{Basic idea}

The new algorithm addresses these issues by explicit use
of the sensitivities, $S_{ij}(\vec \phi) = \frac{\partial p_i}{\partial
  \phi_j}$, of each band $i$ to each AP $j$. These are estimated by fitting a
(smooth) function, ${\hat p_i} = f_i(\vec \phi)$ to each band, which I refer
to as the {\em forward model} (Fig.~\ref{formod}). Sensitivities are estimated from these via first
differences.  These are used to improve the commonly-employed nearest
neighbour (or $\chi^2$ minimization) technique. We locally interpolate the
grid, defining ``optimal'' directions for interpolation using the
sensitivities, and predicting the photon counts at off-grid points using the forward
model.

\section{The algorithm}

\begin{figure}[t]
\epsscale{0.75}
\plotone{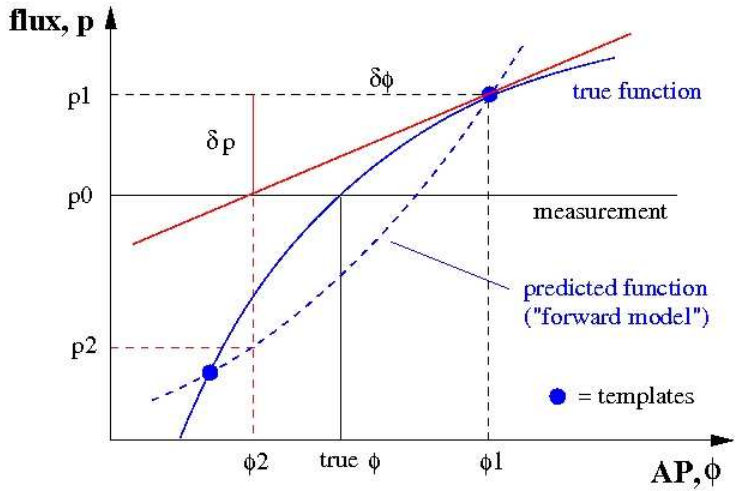}
\caption{Sketch of the local iteration principle (described in the text) for
  $I=1$ and $J=1$.} \label{principle}
\end{figure}

Task: estimate APs of measured vector $\vec p_0$.
The core algorithm (for $I=1, J=1$) is as follows, where subscripts refer here to iterations (see
Fig.~\ref{principle})

\begin{enumerate}
\item Fit the forward model to the grid, $\hat p = f({\vec \phi})$
\item Initialize: find nearest grid neighbour to $\vec p_0$. Call this $({\vec p_1}, {\vec \phi_1})$
\item Use the forward model to calculate the local sensitivities, $\frac{\partial \phi}{\partial p}$
\item Calculate the discrepancy (residual), $\delta p_n = p_n - p_{n-1}$ ($= p_1 - p_0$ for the first iteration)
\item Make a step in AP space,
$ \phi_{n+1} = \phi_n - \left (\frac{\partial \phi}{\partial p}
\right)_{\phi_n} \delta p_n $. This is the new AP prediction
\item Use the forward model to predict the corresponding (off-grid) flux, $p_{n+1}$
\item Iterate steps 3--6
\end{enumerate}

\noindent 
For the general case of $I>1$, step 5 is simply an average over the update
calculate for each band, i.e.\ $\delta \phi = - \sum_i \frac{\partial
  \phi_j}{\partial p_i} \, \delta p_i$. For multiple APs ($J>1$), we can write
this in matrix format as $\delta \vec{\phi} = - R \, \delta {\vec p}$, where
$R = [R_{ji}]$ is the $J \times I$ matrix of reciprocal sensitivities, i.e.\ 
$R_{ji} = S_{ij}^{-1}$ (mathematical rigour being sacrificed to some degree).
The actual algorithm is a bit more complex (e.g.\ modified step size).

\section{Application and results}

\begin{figure}[t]
\epsscale{0.90}
\plotone{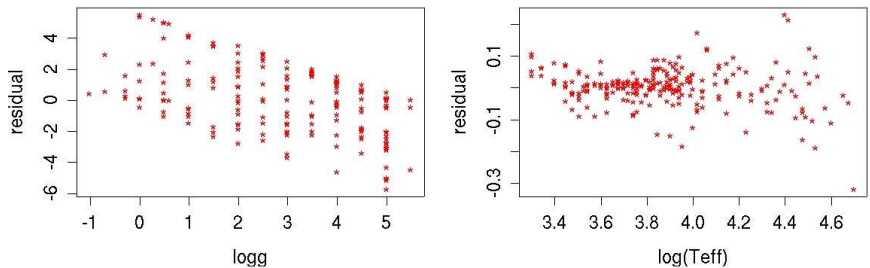}
\caption{Parameter estimation residuals for \logg\ (left) and \logteff\ (right).} \label{residuals}
\end{figure}

I apply the algorithm to a set of synthetic optical photometry of stars
showing variation in \teff\ and \logg. The $I=11$ photometric bands
are part of a system originally designed for Gaia (Jordi et al.\ 2006).  The
data show a large variation in APs (Fig.~\ref{residuals}), and the grid is
quite sparse, comprising just 233 objects. The test set contains 234 objects,
with no AP combinations in common between the grid and test set.  The
signal-to-noise ratio of the test set has been reduced to 10 per band.

The algorithm applied is actually a simpler version of the general case
described above: the forward model in steps 3 and 6 is a local linear
interpolation (i.e.\ a plane) of the neighbours in AP space (not data space!)
to the current point under consideration.\footnote{Neighbours are selected so
  as to ``surround'' the current point in AP space as well as possible,
  providing a sort-of ``bracketing'' (a concept which is only properly defined in one
  dimension).}  Thus the forward models are robust but suboptimal (as we are
no longer taking advantage of the reasonable assumption that $p_i = f_i(\vec
\phi)$ is smooth).

\begin{figure}[t]
\epsscale{0.90}
\plottwo{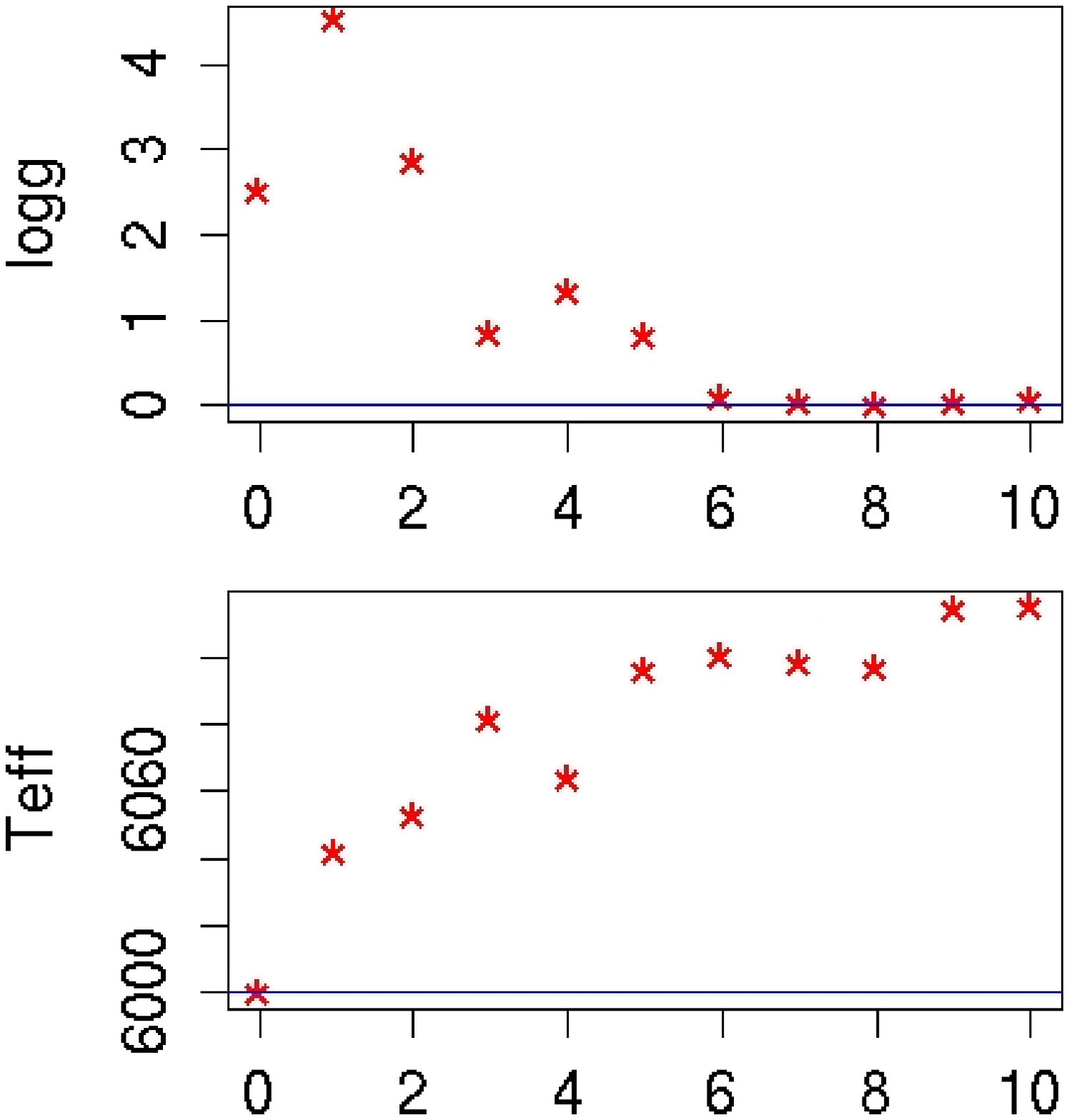}{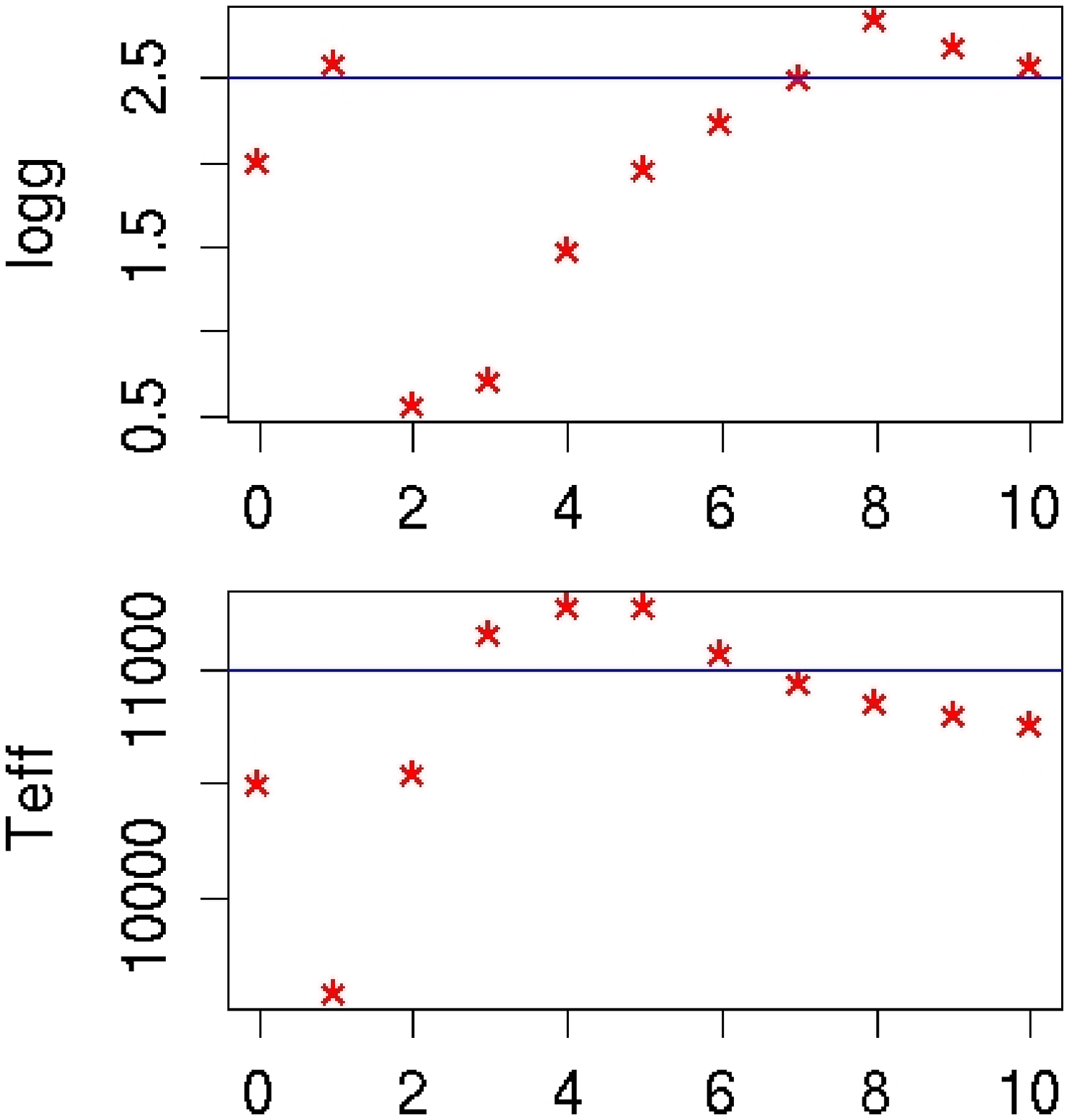}
\caption{AP estimate vs.\ iteration number for for two sources (left and
  right), showing examples of correct and incorrect convergence. The 
  horizontal line shows the true parameters.} \label{iterations}
\end{figure}

The progress of the algorithm in terms of the AP estimates is demonstrated in
Fig.~\ref{iterations} for two of the 234 test cases. These have been selected
to demonstrate both good and poor convergence for the two APs.

The residuals of the estimates over all 234 objects are shown in
Fig.~\ref{residuals}, plotted against the true APs. The RMS errors are
1.65\,dex in \logg\ and 0.042\,dex in \logteff\ (corresponding to 480\,K at
5000\,K). These compare to 0.99\,dex for \logg\ and 0.028\,dex for \logteff\ 
(320\,K at 5000\,k) for an SVM model trained and tested on the same data.
The systematic error in \logg\ is also seen with the SVM and is characteristic
of the weak AP problem (but not, I believe, unassailable).

\section{Conclusions}

The algorithm currently performs slightly worse than one of the best
generic regression algorithms available (an SVM), yet this is not bad
considering that (a) it is in an early stage of development, and (b) the
results were obtained using a (suboptimal) {\em linear} forward model.  Unlike
ANNs, the dimensionality of the algorithm's fitting depends on the number of APs ($J$),
not the number of data dimensions ($I$), so it should scale well to typical
spectral problems ($J$ is a few, $I$ is a few thousand).  Moreover, the method
has the ability to detect and report multiple solutions which arise from
degeneracies in the data (see Fig.~\ref{formod}).

\acknowledgments

I thank C.~Tiede and K.~Smith for useful discussions.

\end{document}